\documentclass[aps,pre,twocolumn,superscriptaddress]{revtex4}
\pdfoutput=1
\usepackage{amssymb}
\usepackage{amsfonts}
\usepackage{rotating}
\usepackage{amsmath}
\usepackage{graphics}
\usepackage{epsfig}
\usepackage{hyperref}

\newcommand{\lb}{\label}
\newcommand{\be}{\begin{equation}}
\newcommand{\ee}{\end{equation}}
\newcommand{\av}[1]{\langle #1 \rangle}
\newcommand{\bea}{\begin{eqnarray}}
\newcommand{\eea}{\end{eqnarray}}

\begin{document}


\title{Contrasting Effects of Strong Ties on SIR and SIS Processes in Temporal Networks}

\author{Kaiyuan Sun}
\affiliation{Laboratory for the Modeling of Biological and Socio-technical Systems, Northeastern University, Boston MA 02115 USA}

\author{Andrea Baronchelli}
\affiliation{Department of Mathematics, City University London, Northampton Square, London EC1V 0HB, UK}

\author{Nicola Perra}
\affiliation{Laboratory for the Modeling of Biological and Socio-technical Systems, Northeastern University, Boston MA 02115 USA}

\date{\today} \widetext

\begin{abstract} 
Most real networks are characterized by connectivity patterns that evolve in time following complex, non-Markovian, dynamics. 
Here we investigate the impact of this ubiquitous feature by studying the Susceptible-Infected-Recovered (SIR) and  Susceptible-Infected-Susceptible (SIS) epidemic models on activity driven networks with and without memory (i.e., Markovian and non-Markovian). We find that memory inhibits the spreading process in SIR models by shifting the epidemic threshold to larger values and reducing the final fraction of recovered nodes. In SIS processes instead, memory reduces the epidemic threshold and, for a wide range of diseases' parameters, increases the fraction of nodes affected by the disease in the endemic state. 
The heterogeneity in tie strengths, and the frequent repetition of strong ties it entails, allows in fact less virulent SIS-like diseases to survive in tightly connected local clusters that serve as reservoir for the virus. We validate this picture by studying both processes on two real temporal networks.\end{abstract}

\maketitle

Virtually any system can be represented as a network whose basic units are described as nodes and its interactions as links between them~\cite{butts09-1,newman10-1,barrat08-1,gcalda}. In general, connections are not static, but evolve in time subject to nontrivial dynamics~\cite{holme11-1}. Consider for example face to face or online interaction networks where individuals talk and exchange information through evolving contacts~\cite{cattuto2010dynamics,Isella:2011,panisson11-1,weng13-1}. 
Recent advances in technology have allowed researchers to collect, monitor and probe such interactions generating an unprecedented amount of time-resolved high resolution data~\cite{vespignani09-1}.  The analysis of such real systems has exposed the limits of canonical static and annealed network representations~\cite{holme11-1} calling for the development of a new theory to understand network's temporal properties.
In particular, the recent data deluge has allowed researchers to start identifying the effects that time varying topologies have on dynamical processes taking place on them~\cite{Isella:2011,panisson11-1,morris93-1,morris07-1,clauset07,alex12-1,Rocha:2010,Stehle:2011nx,Karsai:2011,Miritello:2011,dynnetkaski2011,albert2011sync,Parshani:2010,Bajardi:2011,consensus_temporal_nrets_2012,starnini_rw_temp_nets,pfitzner13-1,karsai13-1,lambiotte12-1,toro2007,perra12-1,ribeiro12-2,perra12-2,liu13-1,starnini13-2}. Prototypical examples are the spreading of memes, ideas, and infectious diseases. 
All of these phenomena can be described as diffusion processes on contact networks and are affected by the ordering, concurrence, duration, and heterogeneity in nodes' activities and connectivity patterns~\cite{Isella:2011,panisson11-1,morris93-1,morris07-1,clauset07,alex12-1,Rocha:2010,Stehle:2011nx,Karsai:2011,Miritello:2011,dynnetkaski2011,albert2011sync,Parshani:2010,Bajardi:2011,consensus_temporal_nrets_2012,starnini_rw_temp_nets,pfitzner13-1,karsai13-1,lambiotte12-1,toro2007,perra12-1,ribeiro12-2,perra12-2,liu13-1,starnini13-2,liu13-2,scholtes13-1,Miritello2011Dynamical}.  

One of most distinctive properties of social networks is the heterogeneity of interaction strength~\cite{granovetter73-1,Wasserman1994Social,dunbar92-1}.  Individuals remember their inner circle of friends and most important connections, activating some links more often than others, thus building up strong and weak ties with their peers. In other words, the creation of links is not a Markov process~\cite{pfitzner13-1,karsai13-1,granovetter73-1,Wasserman1994Social,dunbar92-1,rosvall13-1}.  
While this property has been studied in detail in static networks~\cite{granovetter73-1,Wasserman1994Social,flache96-1,CPLX:CPLX10053,dodds03-1,onnela06-1,Shi200733,Xiang:2010:MRS:1772690.1772790}, its understanding in the context of time-varying graphs is still far from complete. 
Indeed, only a few studies have tackled this subject, and each with  a different approach ~\cite{karsai13-1,scholtes13-1,rosvall13-1,lambiotte13-1}. Nonetheless, these studies have revealed a rich phenomenology. In particular, non-Markovian link dynamics has been shown to be responsible for changing the spreading rate of diffusion processes, either slowing them down or, perhaps surprisingly, speeding them up~\cite{karsai13-1,scholtes13-1,rosvall13-1,onnela06-1,lambiotte13-1,Karsai2011Small,Karsai2012Correlated}. 

Here we study the effects of memory on two different classes of epidemic spreading models, namely the Susceptible Infected Recovered (SIR) and the Susceptible Infected Susceptible (SIS) models~\cite{keeling08-1}. We consider a recently proposed class of time-varying networks called activity driven models~\cite{perra12-1,karsai13-1}, based on the observation that the propensity of nodes to initiate a connection (the activity) is heterogeneously distributed. In its basic formulation node activities are modeled with accuracy but the link creation is assumed to be Markovian. While such an approximation allows analytical treatments~\cite{perra12-1,ribeiro12-2,perra12-2,liu13-1,starnini13-2,liu13-2}, it does not capture many real properties of time-varying networks such as the memory of individuals. Recently, this limitation has been overcome with the introduction of a non-Markovian generalization of the modeling framework based on a simple reinforcement mechanism that allows one to reproduce with accuracy the evolution of individual's contacts~\cite{karsai13-1}.

We study the dynamical properties of SIR and SIS models on activity driven networks with and without memory. 
In particular, we consider one of the most important dynamical properties of epidemic diffusion process, namely the epidemic threshold, defining the conditions necessary for the spreading of the disease to a macroscopic fraction of the population~\cite{keeling08-1}. We also consider the effect of the disease on the population evaluating the final fraction of recovered nodes, in SIR processes, and the fraction of infected nodes in the endemic state, reached above threshold in SIS dynamics.

We find that memory acts in different ways on SIR and SIS models. In SIR processes the epidemic threshold is shifted to larger values, making  the spreading of the disease more difficult. Also, the final fraction of recovered nodes is significantly reduced. In SIS dynamics memory moves the epidemic threshold to smaller values and shifts the endemic state, for a wide range of disease's parameters, to larger values. Thus, non-Markovian dynamics might facilitate the spreading of SIS-like diseases, like sexual transmitted illnesses, that can survive reaching an endemic  state, in tightly connected clusters. The difference between the two models is due to the fundamentally different natures of the two processes that induce distinct behaviors also in the case of static networks~\cite{ferreira12-1,castellano10-1,goltsev12-1}.

Finally, we consider two real-world networks built using messages exchanged between users on Twitter and co-authorships of papers in a scientific journal. To isolate the role of non-Markovian dynamics in this case, we compare the spreading of SIR and SIS processes unfolding on real networks with the same dynamics unfolding on a randomized version of them.  Interestingly, in the case of SIS processes the results are qualitatively similar to what is observed in synthetic networks. In the case of SIR dynamics we do not observe a significant change in the epidemic threshold. However, consistently to what observed in synthetic networks, the real non-Markovian dynamics hampers the disease spreading reducing significantly the final fraction of recovered nodes. 

In this section we describe the modeling framework used to produce the considered synthetic time-varying networks. 

\subsection{Memoryless activity driven models (ML)}

\begin{figure}
\includegraphics[width=\columnwidth]{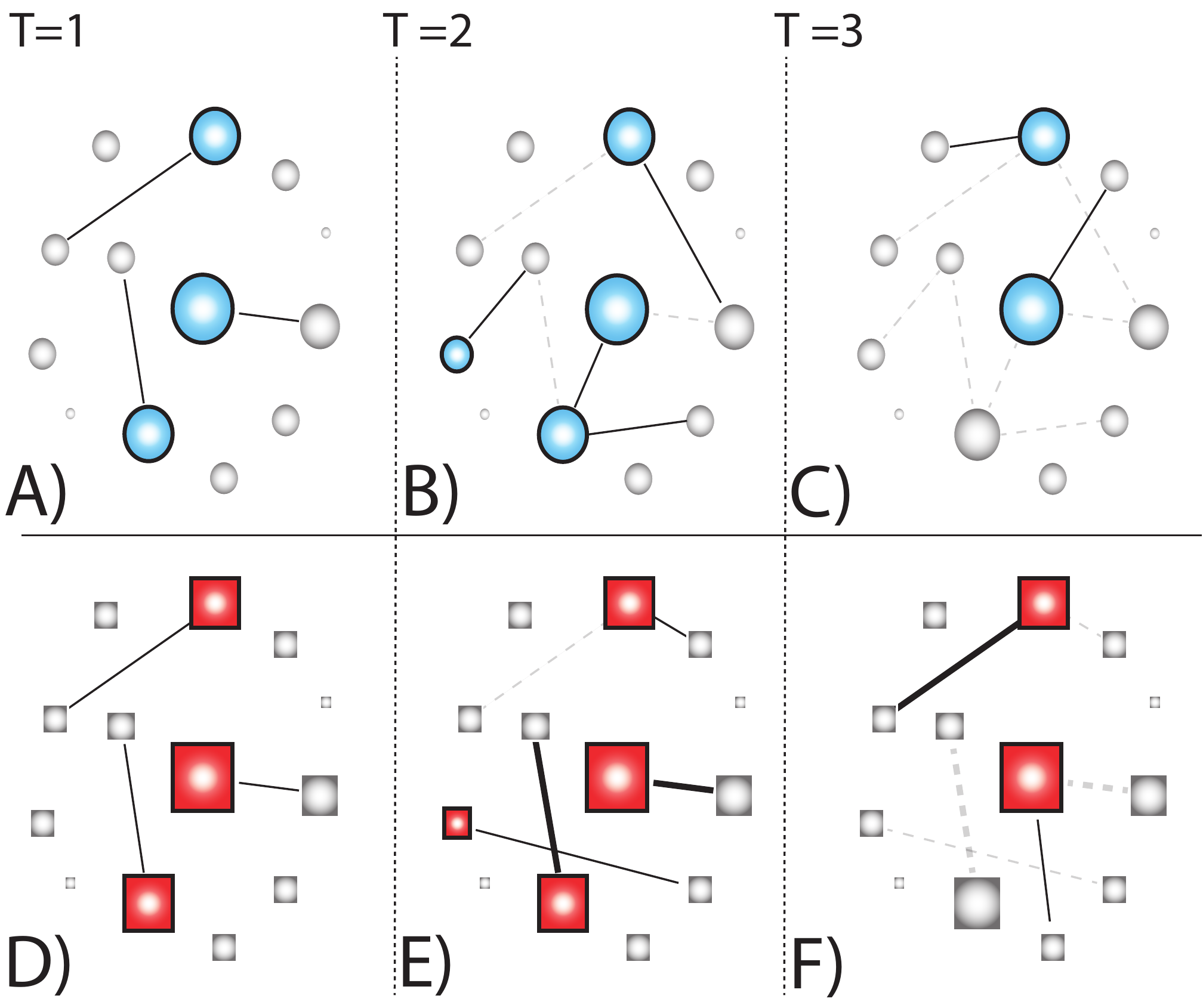}
\caption{ Schematic representation of ML and WM activity driven models. In order to better contrast the two different models we fixed the same activity distribution in both cases, and we show a simulation scenario in which for both models the active nodes at each time step are the same. In grey (dashed lines) we show links previously initiated, while in black (solid lines) links activated in the current time step. Active nodes are shown in light blue for the ML model while in red for the WM, and marked with a tick border.  The size of each node is proportional to the activity, and the thickness of each link describes its weight. Panels A,  B and C show ML networks  at three different time steps $T=1,2,3$. Panels D, E, and F show WM networks at three different time steps $T=1,2,3$ }
\label{fig:Fig1}
\end{figure}

In their basic formulation activity driven models are memoryless.  Each node is characterized by an activity rate $a$, extracted from a distribution $F(a)$, describing its probability per unit time to establish links.  To account for the observation that human behaviors are characterized by broad activity distributions we will consider power-law distributions of activity $F(a)=Ba^{-\gamma}$ ($\epsilon \le a \le 1$), unless specified differently.
In this setting, the generative process of the network is defined
according to the following rules (see Figure~~\ref{fig:Fig1}):

\begin{enumerate}
\item At each discrete time step $t$ the network $G_t$ starts with $N$
  disconnected vertices;
\item With probability $a_i \Delta t$ each vertex $i$ becomes active
  and generates $m$ links that are connected to $m$ other randomly
  selected vertices. Non-active nodes can still receive connections
  from other active vertices;
\item At the next time step $t + \Delta t$, all the edges in the
  network $G_t$ are deleted. 
 \end{enumerate}
Thus, all interactions have a constant duration $\Delta t$, that without loss of generality we fix to one, i.e. $\Delta t=1$. 

At each time step the network $G_t$ is a simple random graph with low average connectivity. Indeed, on average the number of active nodes per time step is $N\langle a \rangle$, corresponding to an average number of edges equal to $mN\langle a \rangle$, and an average degree $\langle k \rangle=2 m \langle a \rangle$. However, integrating the links over $T$ time steps, so that $T/N \ll 1$, induces networks whose degree distribution follows the activity functional form~\cite{perra12-1,starnini13-2} so that, for example, broad distributions of activity will generate broad degree distributions. The
creation of hubs (highly connected nodes) results from the presence of nodes with high activity
rate, which are more prone to repeatedly engage in interactions. \\

\subsection{Activity driven models with memory (WM)}

It has long been acknowledged that links in real-world networks can be grouped in (at least) two classes, namely strong and weak ties~\cite{granovetter73-1,Wasserman1994Social}. The first represent connections that are activated often and describe, for example, the inner social circle of each node. 
The latter describe occasional contacts that are activated sporadically. From a modeling standpoint these different classes of links can be described considering individuals as non-Markovian. Indeed, the evolution of their ego-centered networks is deeply influenced by their social memory. Interestingly, 
empirical observations indicate that  the probability for an individual that had interacted with $n$ people to initiate a connection towards a $n+1$th individual is a function of $n$. More precisely, the analysis of a large-scale mobile phone dataset~\cite{karsai13-1} identified the relation
\be
P_k(n+1)=\frac{c_k}{n+c_k},
\ee
where $k$ is the total number of other nodes contacted measured at the end of the datasets, and $c_k$ is a constant mildly dependent on the degree. Thus, setting for simplicity $c_k=1$ $\forall \,\ k$, it is possible to generalize the activity driven framework accounting for individuals' memory~\cite{karsai13-1}. Given, as for the ML case, $N$ nodes each characterized by an activity rate $a$ extracted from a distribution $F(a)$, the generative process of the WM network is defined according to the following rules  (see Figure~\ref{fig:Fig1}):

\begin{enumerate}
\item At each discrete time step $t$ the network $G_t$ starts with $N$
  disconnected vertices;
\item With probability $a_i \Delta t$ each vertex $i$ becomes active
  and generates $m$ links; 
 \item Each link is established with probability $1/(n_i+1)$ at random, and with probability $n_i/(n_i+1)$ towards one of the $n_i$ 
 previously connected nodes.
 Non-active nodes can still receive connections
  from other active vertices;
\item At the next time step $t + \Delta t$, the memory of each node is updated and all the edges in the
  network $G_t$ are deleted. 
 \end{enumerate}
 
\begin{figure}
\includegraphics[width=\columnwidth]{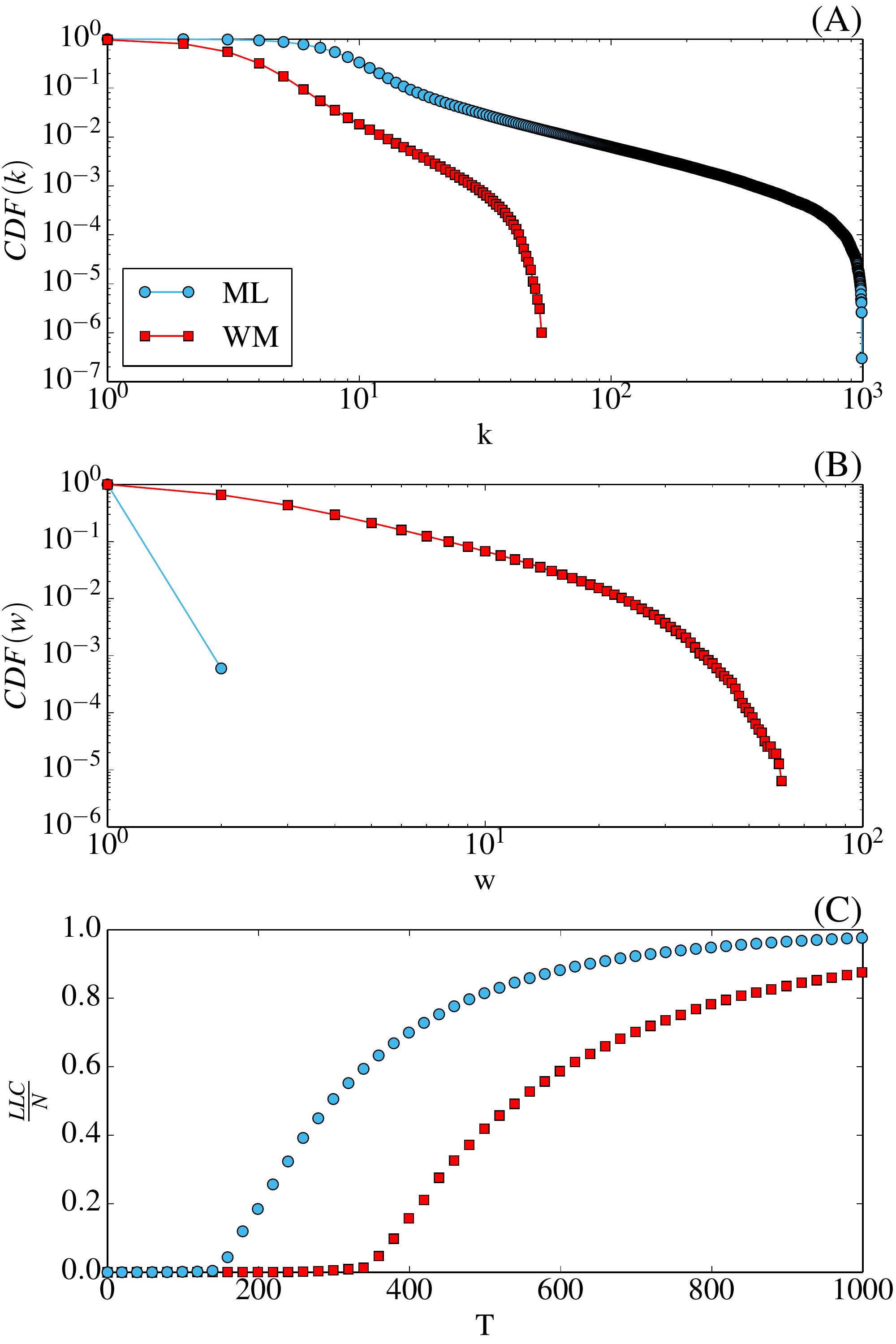}
\caption{The ML and WM activity driven networks. A) Cumulative degree distribution for both ML ( blue circles) and RP (red squares) activity driven networks integrated for $T$ time steps. B) Cumulative weight distribution for the same networks. C) Emergence of the largest connected component (LCC) in ML  and WM activity-driven networks as a function of time. In particular, we plot the normalized size of the LCC, $LCC/N$, as a function of the integrating time $T$. For all the panels we fix $N=10^5$, $m=1$, and $\epsilon=10^{-3}$, $T=10^3$, and consider $10^2$ independent realizations.}
\label{fig:Fig_stru}
\end{figure}

The structural properties of time-aggregated ML and WM activity driven networks are fundamentally different. As is clear from Figure~\ref{fig:Fig_stru} ML networks show a heavy-tailed cumulative degree and a  homogeneous weight distribution, where the weights measuring the number of times each link is activated reflect the Markovian links' creation dynamics (see Figure~\ref{fig:Fig_stru}-B). On the other hand, WM networks show a broad degree distribution, steeper than the one observed in ML systems, (see Figure~\ref{fig:Fig_stru}-A) and a heavy-tailed weight distribution indicating the heterogeneity of tie strengths (see Figure~\ref{fig:Fig_stru}-B). In Figure~\ref{fig:Fig_stru}-C we also compare the behavior of the largest connected component (LCC) integrating the links as a function of time. Interestingly, in ML networks the LCC appears earlier. Memory slows down the growth of the connected component as individuals are more likely to activate previous connections.\\

\section{SIR and SIS models in activity driven networks}
\label{diff_synt}

We consider two classic epidemic models, namely the SIR and SIS model~\cite{keeling08-1}. In both cases the population is divided in compartments indicating the health status of individuals.
In the SIR model nodes can be in the susceptible (S),  infected (I)  or recovered (R) compartments. Susceptible nodes are healthy individuals that never experienced the illness. Infected nodes have contracted the illness and can spread it. Recovered nodes have been cured of the disease and are immune. The model is described by the following reaction scheme:
\be
S+I \xrightarrow{\beta} 2I, \;\ I \xrightarrow{\mu} R.
\ee 
The first transition indicates the contagion process. Susceptible nodes in contact with infected individuals become infected with rate $\beta$. In particular, $\beta$ takes into account the average contacts per node, $\av{k}$, and the per contact probability  of transmission $\lambda$, i.e $\beta=\lambda \av{k}$. The second transition describes the recovery process. Infected individuals recover permanently with rate $\mu$. \\
Whether the disease is able to spread affecting a macroscopic fraction of the network or not depends on the value of the infection rate, the recovery rate and the networks dynamics. In particular, in ML networks the SIR contagion process is able to spread if
\be
\lb{t_sir}
\frac{\beta}{\mu} \ge \xi^{SIR}= \frac{2\av{a}}{\av{a}+\sqrt{\av{a^2}}} .
\ee
See Refs.~\cite{perra12-1,starnini14-1} for the derivation details. The quantity $\xi^{SIR}$ defines the epidemic threshold of the process.  For value of $\beta/\mu<\xi^{SIR}$  the disease will die out. Interestingly, the threshold as a function of the first and second moments of the activity distribution, and completely neglects any time-integrated network representation.   \\
 In the SIS model nodes can be either in the susceptible (S) or infected (I) compartment. The model is described by the following reaction scheme:
\be
S+I \xrightarrow{\beta} 2I, \;\ I \xrightarrow{\mu} S.
\ee 
The first transition is the same as SIR models. In the second transition infected individuals heal spontaneously but instead of becoming immune to the disease move back to the susceptible compartment with rate $\mu$. In ML networks the epidemic threshold of an SIS contagion process, $\xi^{SIS}$, is:
\be
\lb{t_sis}
\frac{\beta}{\mu} \ge \xi^{SIS}= \frac{2\av{a}}{\av{a}+\sqrt{\av{a^2}}},
\ee
\\
See Refs.~\cite{perra12-1,starnini14-1}. Interestingly, the threshold is the same as for the SIR model, i.e. $\xi^{SIS}=\xi^{SIR}$. This is a characteristic of ML activity driven networks and is due to the Markovian link creation dynamics~\cite{perra12-1,liu13-1,starnini14-1}. \\
In this paper we investigate numerically the epidemic dynamics occurring on WM networks. 

\subsubsection*{The SIR process on ML and WM networks}
We consider a SIR model and start the epidemic at $t=0$ with a fraction $I_0=10^{-2}$ randomly selected nodes as seeds. SIR models reach the so called disease-free equilibrium in which the population is divided in:
\be
S_\infty+R_\infty=1,\;\ I_\infty=0.
\ee 
All the variables refer to the density of individuals in the population. The infected individuals will always disappear from the population, as each one of them will eventually recover becoming immune.  Below the threshold, in the thermodynamic limit, $R_\infty \rightarrow 0$. Above the threshold instead $R_\infty$ reaches a macroscopic value, i.e. $R_\infty=\mathcal{O}(1)$. The transition between the two regimes is continuous and the behavior of $R_\infty$ can be studied as a second order phase transition with control parameter $\beta/\mu$~\cite{newman10-1,barrat08-1}. 

\begin{figure}
\includegraphics[width=\columnwidth]{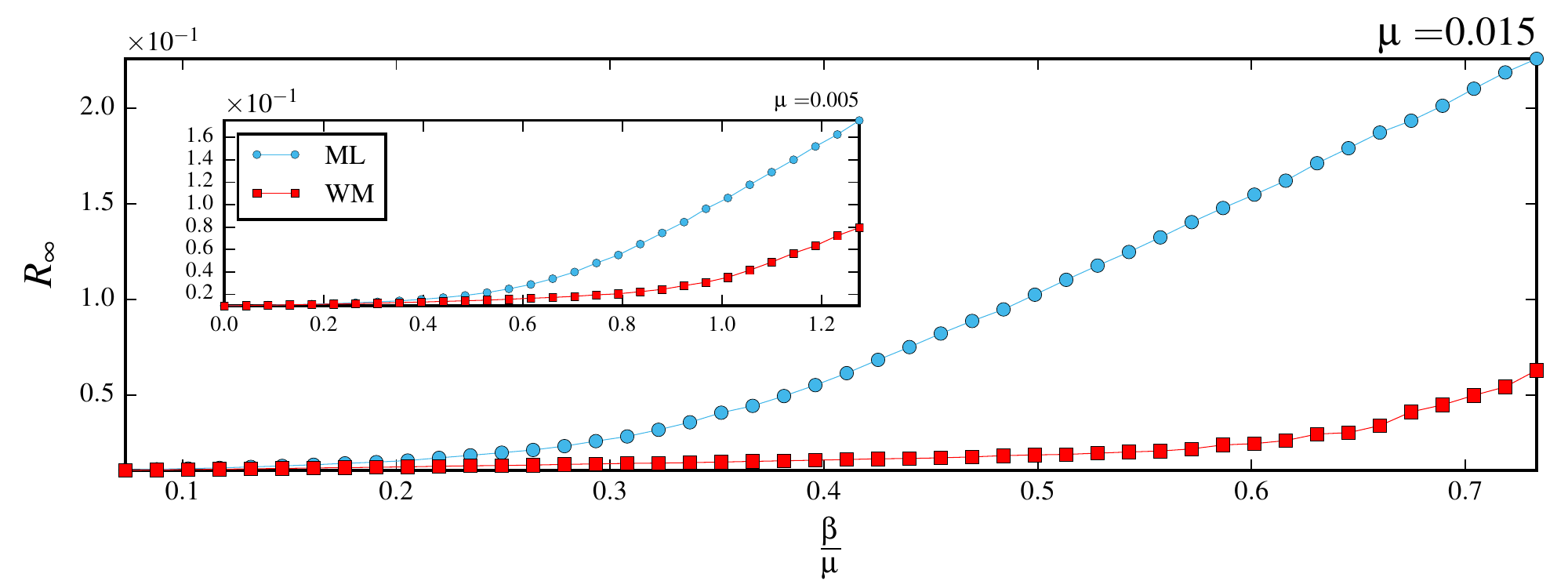}
\caption{SIR spreading in ML (blue circles) and WM (red squares) activity driven networks.  We show $R_\infty$ as a function of $\beta/\mu$.
We fix $N=10^5$, $m=1$, and $\epsilon=10^{-3}$. Each point is evaluated considering $10^2$ independent simulations starting with a fraction of $10^{-2}$ randomly selected nodes. The main plot is done considering $\mu=1.5 \times 10^{-2}$ and the inset $\mu = 5 \times 10^{-3}$. The last point corresponds to $\lambda=1$: the largest value of $\beta/\mu$ for a given network and $\mu$. }
\label{fig:fig3}
\end{figure}

In Figure~\ref{fig:fig3} we show the results obtained by measuring $R_\infty$ in ML and WM networks for different values of $\beta/\mu$. Without loss of generality we fix $\mu=1.5 \times 10^{-2}$ and $\mu=5 \times 10^{-3}$  (inset) and use $\beta$ as free parameter. The epidemic threshold in WM networks is clearly larger than in ML systems. The memory of individuals shifts the threshold to larger values, making the systems less vulnerable to disease spreading. The repetition of interactions within strong ties inhibits the spreading potential of the disease. Indeed, infected individuals will be more likely to contact their inner circle of ties infecting possibly some of them. However, the newly infected nodes will be prone to keep contacting back the initial seeds and eventually recover. 
On the contrary, in ML networks these nodes initiate random connections at each time step increasing their probability of interacting with susceptible individuals. Furthermore, for all the values of $\beta$ sampled, the final fraction of infected nodes in WM networks is significantly reduced. In summary, memory roughly doubles the epidemic threshold of a SIR process and reduces $R_\infty$ making the system more resilient to the spreading. \\

\subsubsection*{The SIS process on ML and WM networks}
\begin{figure}
\includegraphics[width=\columnwidth]{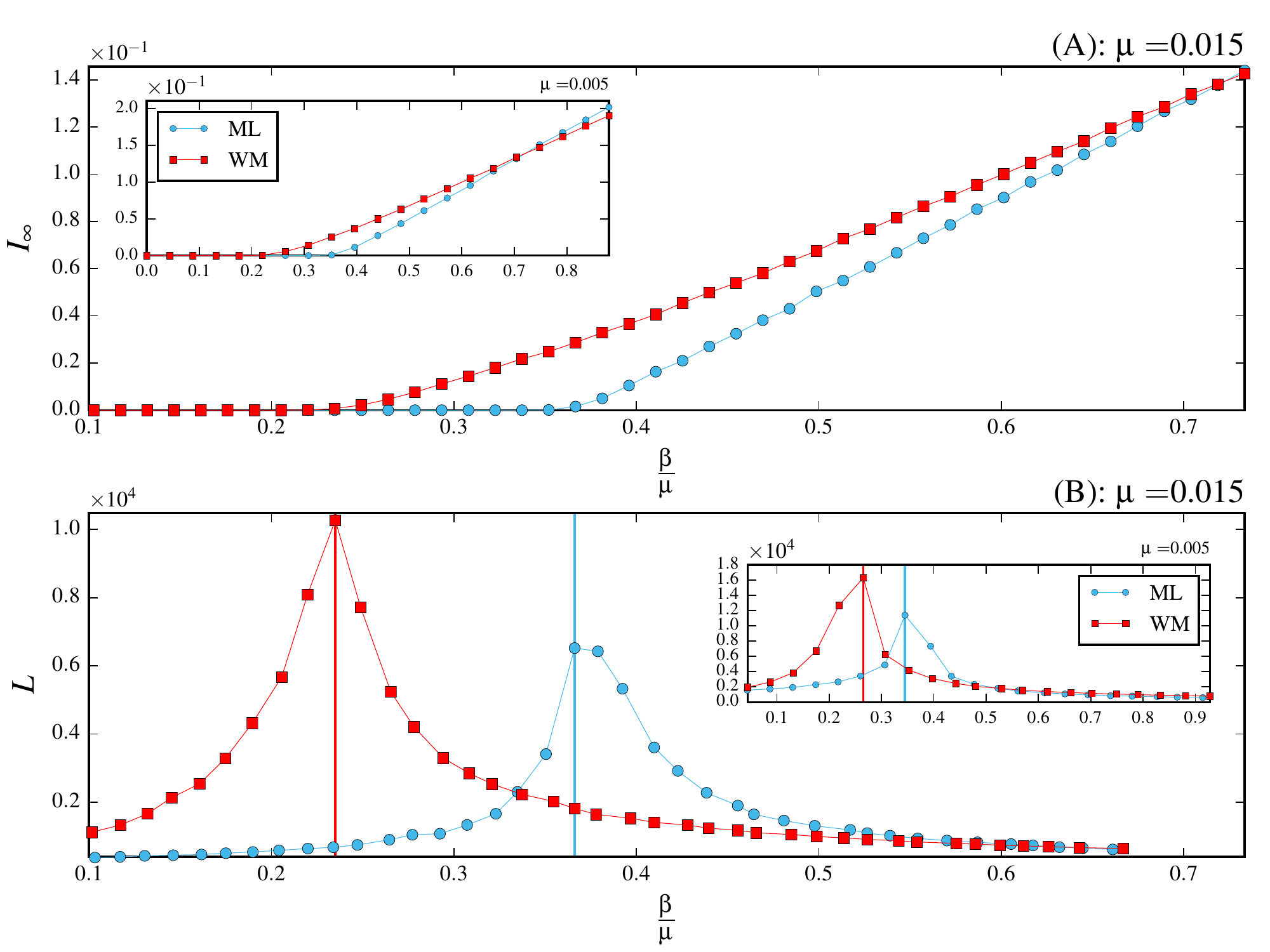}
\caption{SIS spreading in ML (blue circles) and WM (red squares) activity driven networks.  In panel A) we show $I_\infty$ as a function of $\beta/\mu$. In panel  B) instead we plot the life time $L$ as a function of $\beta/\mu$.
We fix $N=10^5$, $m=1$, and $\epsilon=10^{-3}$. Each point is evaluated considering $10^2$ independent simulations starting with a fraction of $10^{-2}$ randomly selected nodes. The main plot is done considering $\mu=1.5 \times 10^{-2}$ and the inset $\mu = 5 \times 10^{-3}$. The last point in each plot corresponds to $\lambda=1$: the largest value of $\beta/\mu$ for a given network and $\mu$. }
\label{fig:fig4}
\end{figure}

We now turn our attention to SIS processes. Also in this case we start the epidemic at $t=0$ with a fraction $I_0=10^{-2}$ of randomly selected nodes as seeds. The nature of this epidemic model is fundamentally different than the SIR. Indeed, above threshold SIS processes show an endemic state characterized by a constant fraction of nodes, $I_{\infty}>0$, in the infected compartment. Below the threshold instead, the process reaches a disease-free equilibrium, i.e. $I_\infty=0$.
In general, in SIS processes the numerical estimation of the threshold is more prone to size and noise effects, due to the subtleties related to the identification of the endemic state
and the fact that $I_\infty$ is not a monotonically increasing quantity as $R_\infty$.  
Therefore, we also consider the life time $L$ and the coverage $C$ of the process as a function of $\beta/\mu$~\cite{boguna13-1}, defining the duration of the process and the fraction of nodes that acquire the infection, respectively. In SIS processes for values of $\beta/\mu$ above threshold  the life time is infinite (endemic state) and the coverage reaches $1$. Below threshold both $L$ and $C$ vanish in the thermodynamic limit.
Interestingly, the life time obtained by averaging over many realizations is equivalent to the susceptibility $\chi$ in standard percolation theory. This method  allows us to detect the threshold precisely~\cite{boguna13-1}.  Indeed, following Ref.~\cite{boguna13-1} we can consider as above threshold any realization that reaches a macroscopic coverage $C$. For small values of the contagion rate the disease dies out quickly and the coverage remains below the threshold $C$, while for very large values of $\beta$ the disease will be able to spread quickly reaching a fraction $C$. For intermediate values of $\beta$, $L$ will increase showing a peak close to the actual epidemic threshold.
Figure~\ref{fig:fig4} shows that the estimation of the threshold performed considering both $I_\infty$ (panel A) and the life time of the process (panel B) using $C=0.5$. We fix $\mu=1.5 \times 10^{-2}$ ( $\mu=5 \times 10^{-3}$ in the inset) and evaluate  $I_\infty$ and $L$ as a function of $\beta$. 

From the two plots we can conclude that the threshold of an SIS process unfolding in WM networks is smaller than in ML systems. This behavior is quite surprising and opposite to what is observed in the case of SIR models. The repeated connections in the ego-centered networks of each node allow the disease to survive in local and small clusters of strong ties making the system more fragile to the disease spreading. Such a behavior is not observed in SIR processes due to the presence of recovered individuals that become immune to the disease and are unable to sustain the spreading with multiple reinfections. Furthermore, in WM networks, for a wide range of $\beta$ values above threshold, $I_\infty$ is shifted to larger values. In this region the disease, due to the repetition of contacts, is able to reach an endemic state that involves a larger fraction of the population. As $\beta$ increases the difference between WM and ML networks reduces and eventually reverses. Indeed, for very large values of the infection rate the disease spreading is favorited by Markovian link dynamics: at each time step active infectious nodes interact with new vertices that, in this regime, can be easily infected.

From this observation we can better understand the effects of memory on the spreading dynamics of SIS processes. The repetition of contacts it entails might counterbalance the effects of small $\beta$ values helping the diffusion. However, for large values of infection rate, memory might hamper the spreading reducing the impact of the disease. In this regime random connections are more efficient. In summary, memory shifts the threshold of SIS processes to smaller values, and for a wide range of infection rates, induces a larger values of $I_\infty$.

\section{SIS and SIR models in real time varying networks}
\label{real}

\begin{figure}
\includegraphics[width=\columnwidth]{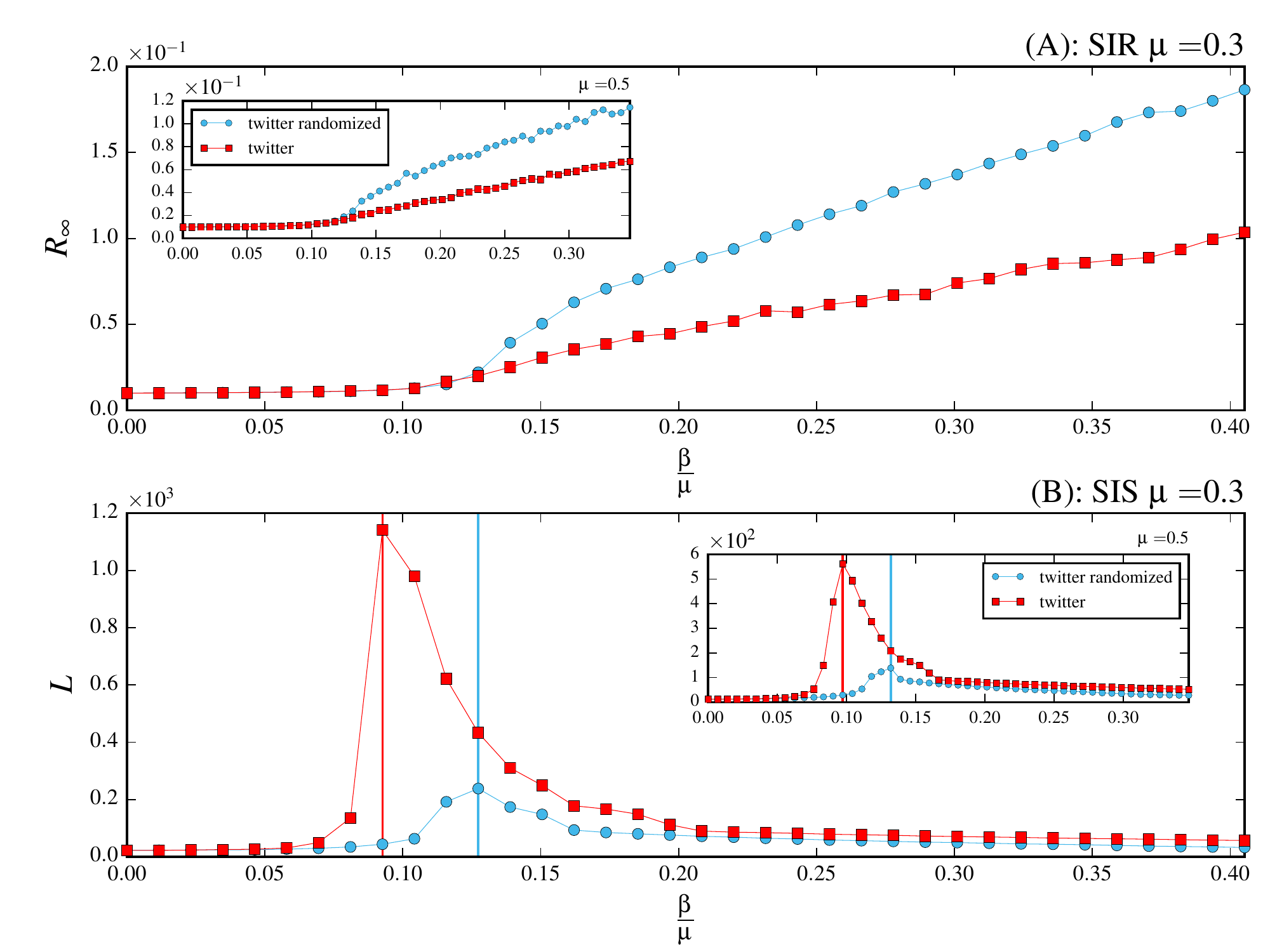}
\caption{SIR and SIS spreading on a real Twitter network (red squares) and on a randomize version of it (blue circles). In panel A) we show the SIR dynamics: $R_\infty$ as function of $\beta/\mu$. In panel B) we show the SIS dynamics plotting the life time $L$ as a function of $\beta/\mu$ Each point is evaluated considering $10^2$ independent simulations starting with a fraction of $10^{-2}$ randomly selected seeds. We set $\mu=0.3$ in main plots and $\mu=0.5$ in the insets.}
\label{fig:fig5}
\end{figure}

In order to validate the results obtained on synthetic time-varying networks we study the dynamical properties of SIR and SIS processes on two real temporal datasets. We consider the interactions between $117436$ Twitter users via $917697$ messages and coarse-grain the data adopting a time resolution of a day. Each user is represented as a node, and at each time step an undirected link is drawn between two nodes if they exchanged at least one message in that time window. The second real dataset is a co-authorship network built considering $268405$ papers published by $55311$ researches in Physical Review Letters (PRL). We adopted the time resolution of one year. Each author is described as a node, and at each time step an undirected link is drawn between two nodes if they co-authored at least one paper in that time window. Arguably such networks are driven by non-Markovian human dynamics as many users and authors tend to interact several times with the same circle of accounts and collaborators. 

In order to single out the effects of memory we consider also two randomized versions of the real networks, where non-Markovian dynamics are washed out. The randomization is performed by reshuffling the interactions for each time stamp, so that memory effects are removed while the sequence of activation times for each node and the degree distribution at each time step are preserved \cite{starnini_rw_temp_nets}.  
In Figure~\ref{fig:fig5}-A we plot the behavior of $R_\infty$ as a function of $\beta/\mu$ for the original Twitter dataset and for the reshuffled version of it considering two values of $\mu$. We do not observe a clear difference between the two epidemiological thresholds. The effects of memory are visible just on the growth of the number of recovered nodes. Indeed, $R_\infty$ increases faster in the randomized network. Thus the repetition of contacts that memory entails hampers SIR spreading processes also on this real network.

\begin{figure}
\includegraphics[width=\columnwidth]{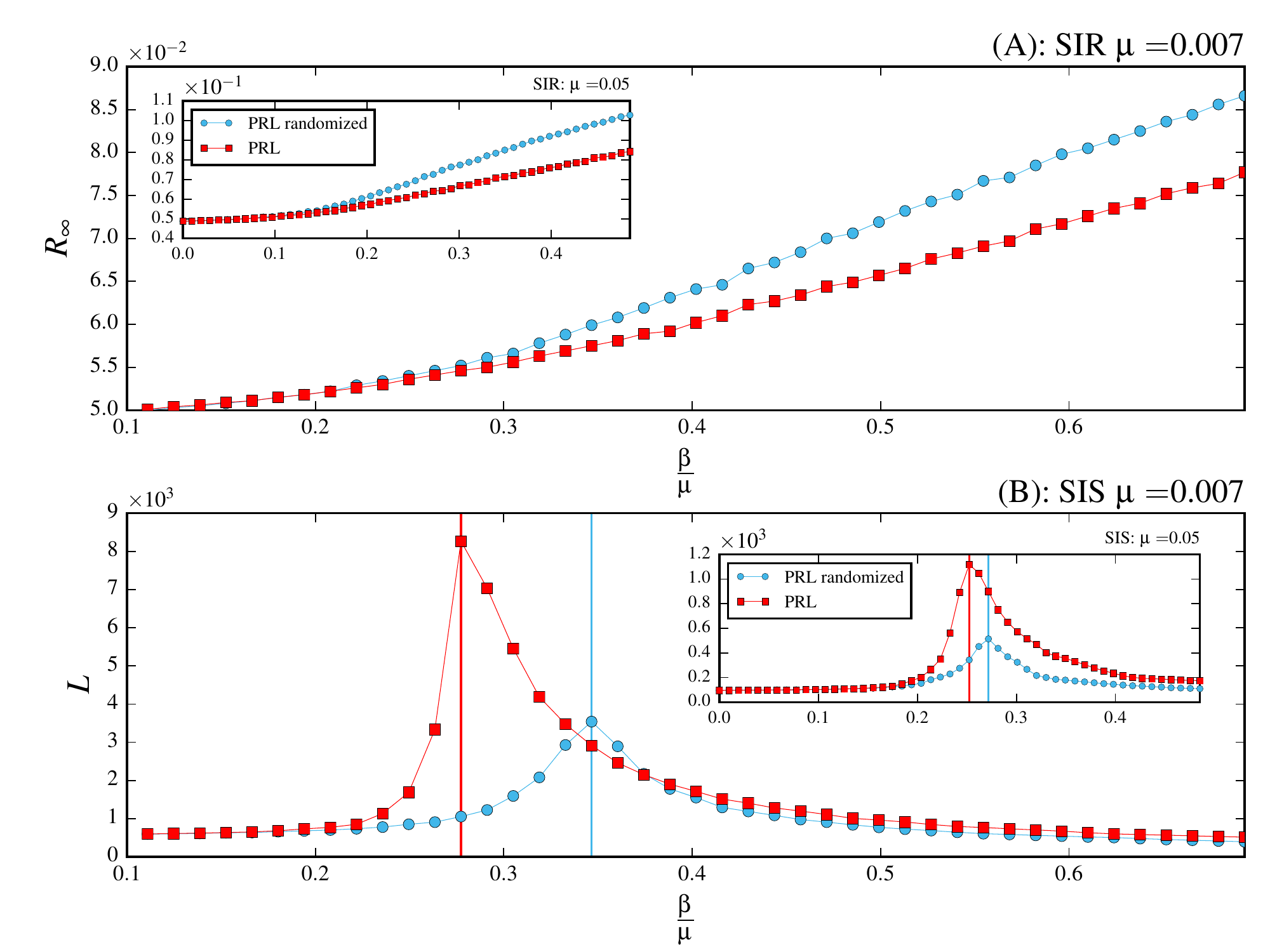}
\caption{SIR and SIS spreading on a real co-authorship network (red squares) and on a randomize version of it (blue circles). In panel A) we show the SIR dynamics: $R_\infty$ as function of $\beta/\mu$. In panel B) we show the SIS dynamics plotting the life time $L$ as a function of $\beta/\mu$ Each point is evaluated considering $10^2$ independent simulations starting with a fraction of $10^{-2}$ randomly selected seeds. We set $\mu=7 \times 10^{-3}$ in main plots and $\mu=5 \times 10^{-2}$ in the insets. }
\label{fig:fig6}
\end{figure}

In Figure~\ref{fig:fig5}-B we plot the behavior of the life time, $L$, of an SIS process in the original Twitter network and in its randomized version considering two values of $\mu$. In this case the threshold in the original network is smaller than in the randomized one, analogously to what is observed in synthetic time-varying networks.
Interestingly, also in real networks memory moves the threshold of SIS processes to smaller values facilitating the survival of the disease.
In Figure~\ref{fig:fig6} we show the results of the same simulations considering the PRL collaboration network. Also in this real dataset memory does not change the epidemic threshold of SIR dynamics acting just reducing the final epidemic size $R_\infty$. Furthermore, in the case of SIS spreading, memory shifts the epidemic thresholds to smaller values. 

Overall, these observations on two real temporal networks confirm qualitatively the picture emerging from synthetic time-varying graphs.

\section{Conclusions}
\label{conclu}

In general, real networks are characterized by temporal and non-Markovian dynamics. 
For example, in social networks, individuals interact more frequently with a small set of strong ties keeping memory of their past connections. While this crucial aspect has been analyzed in detail in static networks' representations, very little attention has been devoted to its characterization on temporal networks. 
Here we studied the dynamical properties of SIR and  SIS models in activity driven networks with and without memory.  In order to single out the effects of non-Markovian dynamics we studied the epidemic threshold in basic activity driven models that by construction are Markovian and memoryless, and in a recent generalization of this modeling framework that explicitly consider non-Markovian link dynamics. We found that memory acts on SIR processes making the system more resilient to the disease spreading. On the contrary, memory acts on SIS processes by lowering the epidemic threshold to smaller values and increasing the fraction of infected nodes in the endemic state (for a wide range of disease's parameters) thus possibly making the systems more prone to the disease invasion. In fact, the heterogeneity in ties' strength induces frequent repetition of contacts that allow the survival of SIS-like diseases  in local groups of tightly connected individuals. The illness reaches its endemic state in small clusters that act as reservoir for the virus. \\
Although activity driven models with memory capture fundamental aspects of real time varying networks, they do not account for other important features as appearance of new nodes, disappearance of old ones, and bursty behaviors just to name a few. While the introduction of these ingredients in the modeling framework is left for future work, here we validated the picture obtained from synthetic networks by considering two real time-varying systems, namely the network of communications in Twitter, and a co-authorship network. Interestingly, the results obtained in this case confirm qualitatively the findings observed in activity driven networks for SIS dynamics. In the case of SIR spreading memory does not change the threshold. However, it reduces significantly the final fraction of nodes affected by the disease thus hampering its spread.   \\
In conclusion, the results here presented show that memory can have opposite effects on different classes of spreading processes, and corroborate the important role played by non-Markovian dynamics on the dynamical processes unfolding on temporal networks \cite{karsai13-1,scholtes13-1,rosvall13-1,lambiotte13-1}. 

\section*{Acknowledgments}

The authors are grateful to Alessandro Vespignani for helpful discussions, insights, and comments.

\end{document}